\documentclass[aps,prd,nofootinbib,
floatfix]{revtex4}
\usepackage{amsmath}
\usepackage{amssymb}
\usepackage{graphicx}
\usepackage{rotating}
\usepackage{bm}% bold math
\usepackage{amsmath}

\newcommand{\be}{\begin{equation}}
\newcommand{\ee}{\end{equation}}
\newcommand{\bea}{\begin{eqnarray}}
\newcommand{\eea}{\end{eqnarray}}
\newcommand{\ba}{\begin{array}}
\newcommand{\ea}{\end{array}}

\newcommand{\AddrAHEP}{
  {\it AHEP Group, Instituto de F\'{\i}sica Corpuscular --
    C.S.I.C./Universitat de Val{\`e}ncia \\
    Edificio de Institutos de Paterna, Apartado 22085,
  E--46071 Val{\`e}ncia, Spain}}
\newcommand{\romatre}{
  {\it Dipartimento di Fisica "E. Amaldi" \\
    Universit\'a degli Studi Roma Tre, Via della Vasca Navale 84, 00146 Roma, Italy}}
\begin{document}

\begin{flushright}
IFIC/12-20\\
RM3-TH/12-3
\end{flushright}

\title{Predicting leptonic CP violation in the light of  Daya Bay result on $\theta_{13}$}
\author{D. Meloni} \email{meloni@fis.uniroma3.it} \affiliation{\romatre }
\author{S. Morisi} \email{morisi@ific.uv.es} \affiliation{\AddrAHEP}
\author{E. Peinado} \email{eduardo@ific.uv.es} \affiliation{\AddrAHEP}

\begin{abstract}
In the light of the recent Daya Bay result $\theta_{13}^{DB}=8.8^\circ\pm 0.8^\circ$, we reconsider 
the model presented in \cite{Meloni:2010aw} showing that, 
when all neutrino oscillation parameters are taken at their best fit values of {\it Schwetz et al.} \cite{data1} and 
$\theta_{13}=\theta_{13}^{DB}$, the predicted values of the CP phase are 
$\delta \approx \pm \pi/4$.

\end{abstract}

\maketitle
The Daya Bay Collaboration has recently contributed to the longstanding question of the magnitude
of  $\theta_{13}$ releasing their data on the $\bar \nu_e \to \bar \nu_e$ oscillation \cite{db};
they provide a clear evidence of more than 5 $\sigma$ deviation from zero of the reactor angle:
\begin{eqnarray}
\label{res}
 \sin^2 2\theta_{13} = 0.092 \pm 0.016 \pm 0.005\,.
\end{eqnarray}
Evidence of non-zero reactor angle yields to a potentially measurable CP phase $\delta$ in future neutrino oscillation experiments \cite{Choubey:2010zz}. 
For this reason, it is important to study the possible predictions for the Dirac phase allowed by flavour models.
The purpose of the present letter is to revise the values of the leptonic CP phase $\delta$ implied by the previous result,
in the model proposed in \cite{Meloni:2010aw} and based on the non-abelian discrete symmetry $S_3$. 
The model gives a Fritzsch-like 
texture for the Majorana neutrinos and, as a consequence of the zeros in the neutrino mass matrix and 
the Daya Bay result, predicts peculiar values for the CP phase $|\delta|\approx \pi/4$, contrary to the expectations of many other models which 
give vanishing or maximal $\delta$ (see \cite{xing} for the implications of $\delta = \pm 90^\circ$ on 
the $\mu-\tau$ symmetry after the Daya Bay results).

In the quark sector, Fritzsch-like 
textures \cite{Fritzsch:1979zq} for both the up and down quark mass matrices of the form \cite{others}
\begin{equation}\label{fr} 
M=\left(
\begin{array}{ccc}
0 &A&0 \\
A^* &C&B \\
0 &B^*&D \\
\end{array}
\right)
\end{equation} 
give the well known relation $\tan\theta_{12}=\sqrt{m_1/m_2}$,
which predicts the small Cabibbo angle as a consequence of the strong hierarchy in the masses. 
A texture as in eq.(\ref{fr}) can also be employed for the Majorana neutrino mass matrix;
this is a particular case of the class of two-zero texture \cite{2zeros} which, 
together with the two relations $\Delta m_{atm}=m_3^2-m_1^2$ and $\Delta m_{sol}=m_2^2-m_1^2$, fixes the absolute neutrino mass scale as suggested in \cite{Fritzsch:2009vy}. 
Unlike the quark sector, 
the solar and atmospheric angles can be large due to the fact that in the neutrino sector the hierarchy is not so pronounced. 

Although a vast class of Fritzsch-like textures (and their phenomenological consequences) 
has been already studied in the literature, in the paper of Ref. \cite{Meloni:2010aw}  we have proposed a leptonic model based on 
the permutation symmetry $S_3$ which naturally gives rise to a Fritzsch-type neutrino Majorana mass matrix 
(and, in addition, to a  nearly diagonal charged leptons). 
In the neutrino sector, the Majorana mass matrix is generated by dimension five \cite{Weinberg:1979sa} and six operators
and we obtained a well defined relation between the Dirac phase $\delta$ and the reactor angle $\theta_{13}$ that 
we want to revised here in the light of eq.(\ref{res}).

The  model was based on $S_3$, the group of permutations of three objects,  which  is the smallest  non-Abelian discrete group. 
$S_3$ contains one doublet irreducible representation and two singlets. 
Here we report only the basic ingredients of our model and we remand the interested readers to the original paper \cite{Meloni:2010aw} for
further details.

The Higgs sector was extended from one $SU(2)_L$-doublet to two $SU(2)_L$-doublets, $H_{D}=(H_1,H_2)$ belonging to a doublet 
irreducible representation of $S_3$ and other two $SU(2)_L$ doublets, $H_S$ and $H_S^\prime$, belonging to singlet representations of $S_3$. We also introduced an electroweak scalar singlet $\chi$ 
which turns out to be relevant to give a non-vanishing electron and muon masses. 
In order to have nearly diagonal charged  lepton mass matrix we assumed
two further parity symmetries, so that the global discrete symmetry group of the model is 
$G=S_3 \otimes Z_5 \otimes Z_2$. 
The matter assignment  under $G\otimes SM$ is summarized in Tab.\ref{tab:Multiplet1}.
\begin{table}[h!]
\begin{center}
\begin{tabular}{|c||cccc||ccc|c|}
\hline
fields & $L_D=L_{1,2}$&$L_{3}$ & $l_{R_D}=l_{R_{1,2}}$& $l_{R_{3}}$& $H_{D}$&$H_S$  &$H_S'$& $\chi$ \\
\hline
$SU_L(2)$ & 2& 2& 1&1&2&2&2&1\\
$S_3$  &  $2$& $1$ & $2$&$1$& $2$&$1$ &$1$ &$1$ \\
$Z_2$ &$+$ &$-$ &$+$&$-$ &$+$ &$+$&$-$&$+$\\
$Z_5$ &$\omega^2$ &$\omega$ &$\omega$&$\omega^2$ &$\omega^3$ &$\omega^4$&$\omega^4$&$\omega^2$ \\
\hline
\end{tabular}
\caption{\it Matter assignment of the model of Ref.\cite{Meloni:2010aw}. 
}
\label{tab:Multiplet1}
\end{center}
\end{table}
We have shown that, from the minimization of the potential, a possible solution is given by:
\begin{equation}\label{vev}
\langle H_D \rangle=(v,0) \,.
\end{equation}
After spontaneously symmetry breaking, the charged lepton mass matrix is: 
\begin{equation} 
M_l=\left(
\begin{array}{ccc}
\frac{y_2}{\Lambda} v_S v_\chi& 0&0\\
\frac{y_1}{\Lambda} v v_\chi&\frac{y_2}{\Lambda} v_S v_\chi&0\\
0&0&y_3 v_S\\
\end{array}
\right),
\label{chargedlept}
\end{equation} 
where $v_S =\langle H_S \rangle$,$v_\chi =\langle\chi\rangle$.
When $v_\chi$ is equal to zero only the $\tau$ lepton is massive. The electron and muon masses are generated
by the vev of the scalar $\chi$ and are then suppressed by the large scale $\Lambda$. 
The mass matrix for the charged leptons can be written in terms of the physical lepton masses as:
\be
\label{emmeelle}
M_l=
\left(
\ba{ccc}
\sqrt{m_e m_\mu}& 0 & 0 \\
-m_\mu (1-\frac{m_e}{m_\mu}) & \sqrt{m_e m_\mu} & 0 \\
0 & 0 & m_\tau
\ea
\right),
\ee
and the squared matrix $M_l M^\dagger_l$ is then diagonalized by:
\be
U_L=\left(\ba{ccc}
\frac{1}{\sqrt{1+\frac{m_e}{m_\mu}}} & -\sqrt{\frac{m_e}{m_\mu}}\frac{1}{\sqrt{1+\frac{m_e}{m_\mu}}} &0 \\
\sqrt{\frac{m_e}{m_\mu}}\frac{1}{\sqrt{1+\frac{m_e}{m_\mu}}} & \frac{1}{\sqrt{1+\frac{m_e}{m_\mu}}} & 0 \\
0 & 0 & 1
\ea\right) \approx
\left(
\ba{ccc}
1 & -0.07 &0 \\
0.07 & 1 & 0 \\
0 & 0 & 1
\ea\right)\,.\label{ul}
\ee
The neutrino masses are generated by non-renormalizable operators of dimension 5 and 6  invariant under the group 
$G\times SM$.
The neutrino mass matrix given in Ref.\cite{Meloni:2010aw} is as follows:
\begin{equation}\label{Mnu} 
M_\nu=\left(
\begin{array}{ccc}
0 & 2\,y^\nu_6 (v_S^2+v_S'^2)v_\chi/\Lambda &0 \\
 2\,y^\nu_6 (v_S^2+v_S'^2)v_\chi/\Lambda&(y^\nu_2+y^\nu_3+y^\nu_4) v^2 & y^\nu_9 v v_S' \\
0&y^\nu_9 v v_S'& y^\nu_8 (v_S^2+v_S'^2)\\
\end{array}
\right)\equiv
\left(
\begin{array}{ccc}
0 &b&0 \\
 b&a&c\\
0&c&d\\
\end{array}
\right)\,,
\end{equation}
where $v_S'=\langle H_S'\rangle$.
The mass matrix in eq.(\ref{Mnu}) depends on five real parameters, one of which is related to 
the Dirac phase.
The other four parameters  can be fixed using the experimental information from both 
solar and atmospheric sectors, namely the solar and atmospheric 
mixing angles and squared mass differences. 
The model allows for correlations among the angle $\theta_{13}$ and the CP phase 
$\delta$ that can be easily obtained using  
the zeros of the Fritzsch texture \cite{2zeros}.

\begin{center} 
\begin{figure}[h!] 
\includegraphics[width=8.5cm]{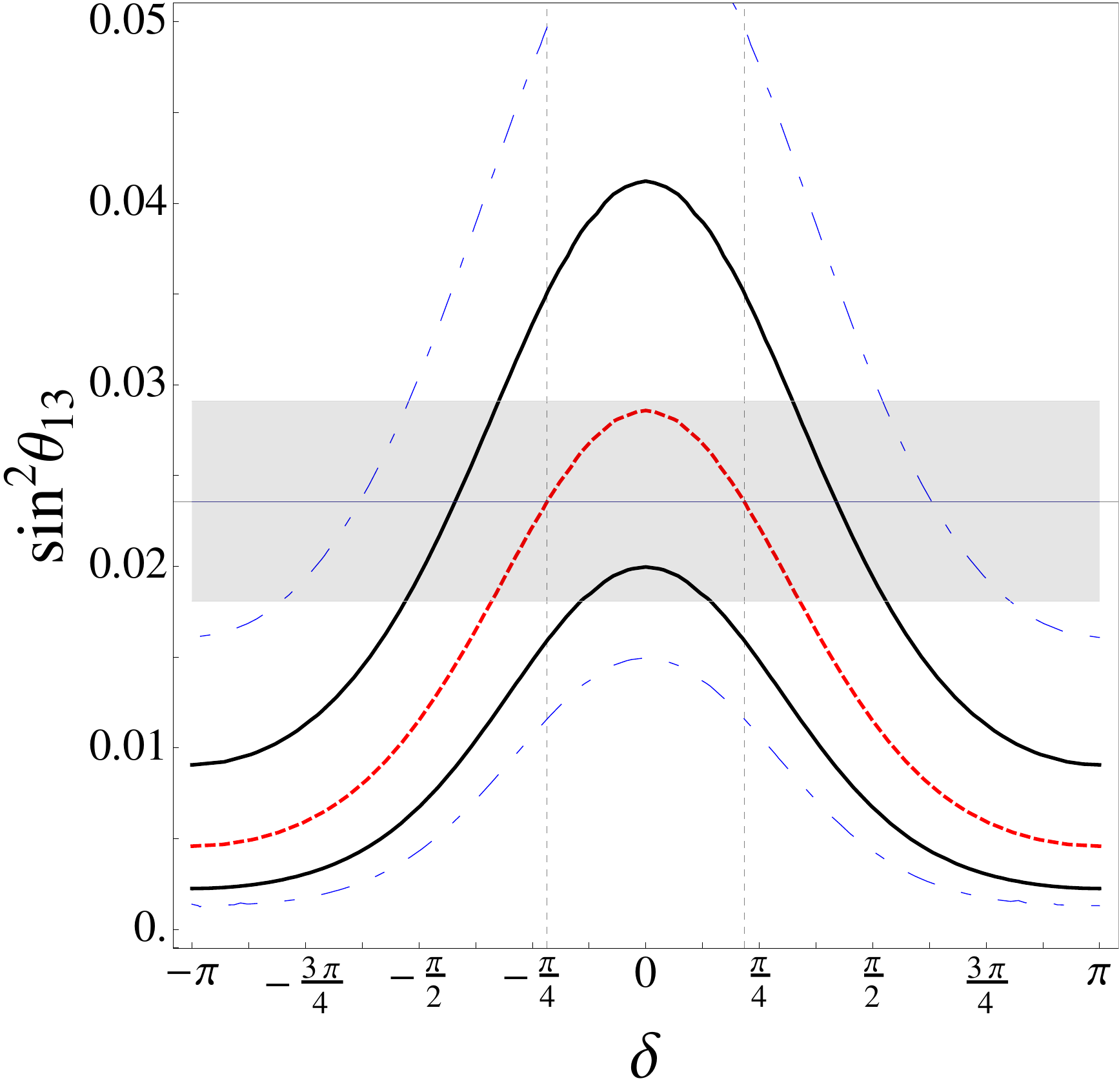}\qquad 
\includegraphics[width=8.5cm]{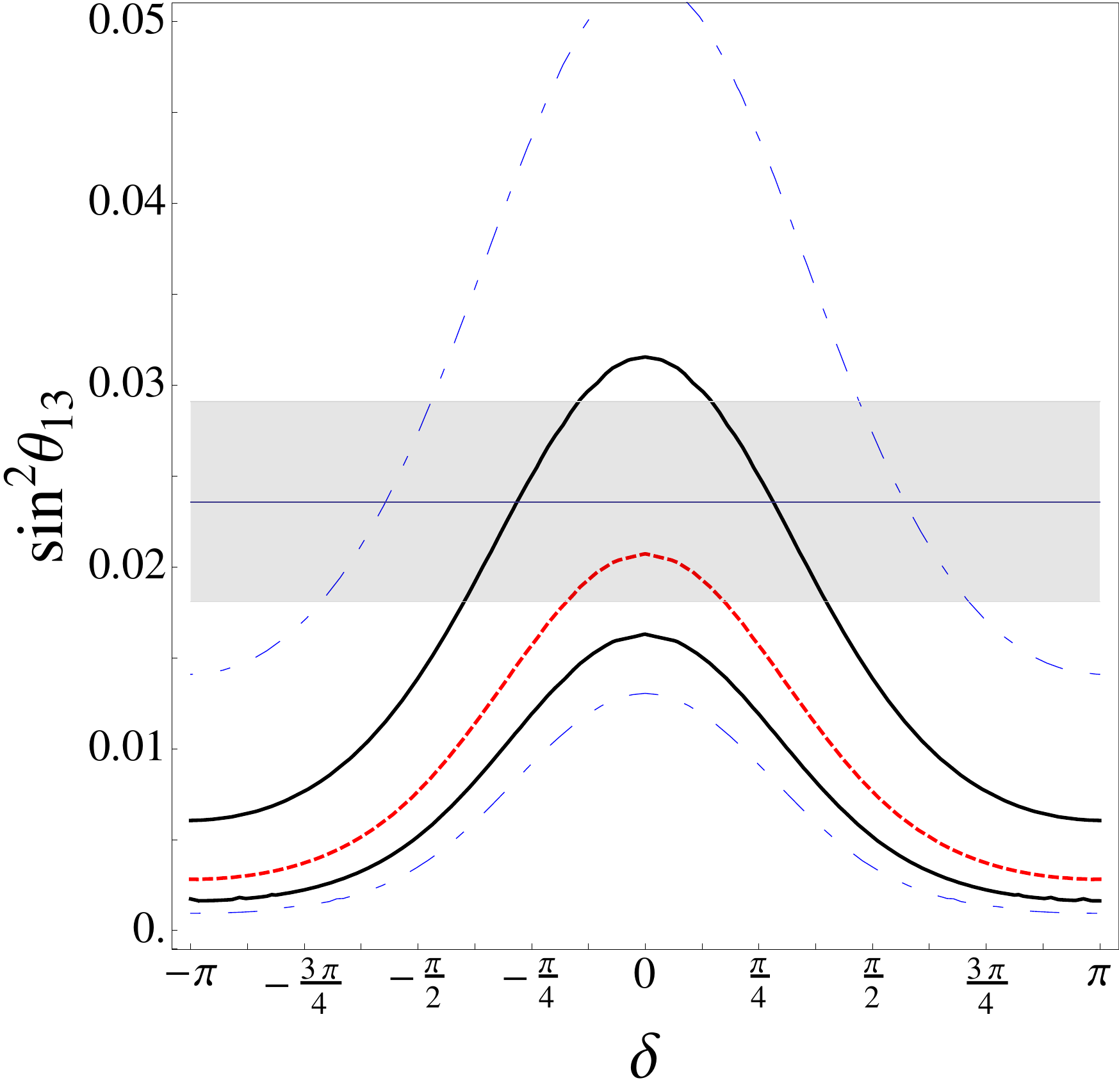}
\caption{\it Correlation among $\delta$ and $\sin^2 \theta_{13}$ as obtained in our model, for two 
different fits, Ref.\cite{data1} (left panel) and Ref.\cite{Fogli:2011qn} (right panel).
The $1\sigma$ result, obtained  varying the other oscillation parameters also in their $1\sigma$ allowed ranges,
is shown with solid lines, whereas the $2\sigma$ result 
is shown with the dot-dashed line. The dashed line is the relation obtained when  
$\theta_{12}, \theta_{23}$ and $\alpha$ are fixed to their best fit values. The horizontal band represents 
% the upper limits on $\sin^2 \theta_{13}$ (upper dashed line) 
the 1$\sigma$ result from Daya Bay\,\cite{db}. 
}
\label{t13-delta}
\end{figure}
\end{center}

In Fig.\ref{t13-delta} we show the dependence of $\sin^2 \theta_{13}$ as a function of $\delta$ as predicted by our model.
The solid line represents the $1\sigma$ correlation when also the other parameters ($\theta_{12}, \theta_{23}$ and the solar-to-atmospheric mass differences ratio $\alpha$) 
are left free to vary in their $1\sigma$ allowed ranges, whereas the $2\sigma$ correlation 
is represented by the dot-dashed line. The dashed line shows the relation between $\theta_{13}$ and $\delta$ when  
$\theta_{12}, \theta_{23}$ and $\alpha$ are fixed to their best fit values. 
The result in eq.(\ref{res}) is enclosed in the horizontal band.
For the sake of completeness, we consider two different fits, the one 
quoted in \cite{data1} (left panel), giving:
\bea
7.41 \times 10^{-5}\,{\text eV}^2<\Delta m^2_{sol}< 7.79\times 10^{-5}\,{\text eV}^2 &\qquad& 2.34 \times 10^{-3}\,{\text eV}^2<\Delta m^2_{atm}< 2.59\times 10^{-3}\,{\text eV}^2
\nonumber 
\\ && \label{tortola}\\
0.298<\sin^2\theta_{12}< 0.329&\qquad& 0.45<\sin^2\theta_{23}<0.58\nonumber
\eea
and the one in \cite{Fogli:2011qn}:
\bea
7.32 \times 10^{-5}\,{\text eV}^2<\Delta m^2_{sol}< 7.80\times 10^{-5}\,{\text eV}^2 &\qquad& 2.26 \times 10^{-3}\,{\text eV}^2<\Delta m^2_{atm}< 2.47\times 10^{-3}\,{\text eV}^2
\nonumber 
\\ && \\
0.291<\sin^2\theta_{12}< 0.324&\qquad& 0.39<\sin^2\theta_{23}<0.5\nonumber\,.
\eea
Let us comment first the results in the left panel.
We observe that, even considering the $2\sigma$ uncertainty, the predicted values for  
$\sin^2 \theta_{13}$ are different from zero so that, to a very good accuracy, our model is compatible with
deviation from $\theta_{13}=0$ for any value of the CP violating phase. 
More remarkably, taking $\sin^2 \theta_{13}\sim 0.024$ as indicated by eq.(\ref{res}),
and all other oscillation parameters to their best fit values quoted in \cite{data1}, 
we predict the CP phase to be $|\delta|\approx \pi/4$. If, instead, we consider the 1$\sigma$ ranges 
on the solar and atmospheric parameters (solid line) we get a CP phase only marginally compatible with 
$\pi/2$, that is:
\bea
-0.53 ~\pi<\delta<0.53~\pi.
\eea
The situation is quite different in the right panel, obtained using the results in \cite{Fogli:2011qn}. In fact,
the dashed line does not intersect the Daya Bay result, as a consequence of a much smaller atmospheric angle
compared to the one quoted in \cite{data1}. This implies a more restricted region for $\delta$, not 
compatible with maximal CP violation:
\bea
-0.40~\pi<\delta<0.40~\pi.
\eea

In this paper we have revised the prediction of the model presented in \cite{Meloni:2010aw} for the leptonic CP phase $\delta$ 
after the recent result on $\theta_{13}$ given by the Daya Bay experiment.
A strong correlation among these two variables is a consequence of 
the two-zero Fritzsch-texture for the neutrino mass matrix; we have shown that, considering 
updated values for the solar and atmospheric oscillation parameters, 
our model predicts a CP phase generally not compatible (or only marginally compatible)
with maximal CP violation. In particular, we get
$|\delta| \approx \pi/4$ if $\theta_{13}$ is taken at the value
indicated by  Daya Bay and all other mixing parameters as in eq.(\ref{tortola}).

\section{Acknowledgments}

We thanks M.Hirsch for the useful discussions. Work of SM and EP supported by the Spanish MEC under grants FPA2011-22975 and MULTIDARK CSD2009-00064
(Consolider-Ingenio 2010 Programme), by Prometeo/2009/091 (Generalitat Valenciana), by the EU ITN UNILHC
PITN-GA-2009-237920. S. M. is supported by a Juan de la Cierva contract. E. P. is supported by CONACyT
(Mexico).
D.M. acknowledges  MIUR (Italy) for financial support
under the program "Futuro in Ricerca 2010 (RBFR10O36O)".

\end{document}